# Flexure based fibre positioners: design optimisation to follow arbitrary focal plane curvature

T Louth[a], S Watson[a], D Montgomery[a]

[a]UK Astronomy Technology Centre, Royal Observatory, Blackford Hill, Edinburgh UK, EH9 3HJ

## ABSTRACT

The use of flexures to achieve fibre positioner motion is being actively investigated by several institutes, for example at the UK Astronomy Technology Centre (UKATC) and Leibniz-Institute for Astrophysics Potsdam [1]. One challenge when designing with flexures is the large number of degrees of freedom available which makes it difficult or impossible to optimise their motion by hand.

In this paper we demonstrate two approaches for optimising flexure geometry to follow arbitrary focal surface curvature and to orient the optical fibre with arbitrary tilt. These approaches are: analytical using MATLAB models and FEA based using Ansys. The approaches are complementary allowing the designer to efficiently explore the parameter space and then do precise optimisation of the flexure geometry. We demonstrate the applicability both to the UKATC's preferred design for WST, and to flexure-based fibre positioner designs generally. We also present a sensitivity analysis relating small changes in design parameters to changes in fibre tip motion. Finally we briefly present the UKATC's preferred geometry for the WST fibre positioner.

**Keywords:** Multi-object spectroscopy, fibre positioners, design optimization, flexure, flexure motion

## 1. INTRODUCTION

**Motivation**

Multi-Object Spectroscopy surveys are an essential tool for gathering astronomical data at sufficiently large volumes to draw statistically significant conclusions. A key enabling technology of MOS instrumentation is the fibre positioners which move optical fibres around the focal plane. There are several established technologies for these, but the next generation of instruments bring new challenges: miniaturisation, simpler manufacture, increased reach for a given footprint and improved speed of focal plane reconfiguration.

Additionally the proposed large survey telescopes such as the Wide-Field Spectroscopic Telescope (WST) are pushing the boundaries of optics manufacturing, which has the potential to result in non-planar and non-telecentric focal surfaces [2]. Maximising light throughput requires the fibre positioners to follow this focal surface curvature within small tolerances, while orienting the fibre tip towards the incoming light.

**Problem Statement**

Flexure based designs offer opportunities for miniaturisation and medium volume manufacture as needed for proposed instruments. The large number of degrees of freedom available in such a design allow for considerable design freedom in the resulting motion, but make optimisation by hand challenging. Flexures are also nonlinear and so cannot be as easily modelled as linkage based designs such as the one presented by this paper's authors at SPIE 2022 [3].

The design that motivated this paper has five principal flexures, the longest of which are each decomposed into three regions: thin "hinges" at each end and a stiffer section in the middle to prevent buckling. Each of the resulting 13 flexure elements can vary in thickness and length, and the five principal flexures can vary in angle with respect to the rotation axis; this totals 31 degrees of freedom. For the purpose of exploring the optimisation approaches in this paper a simpler parallelogram flexure is considered. Four degrees of freedom are independently controlled. The applicability of the approaches to the full flexure is shown in chapter 5.

*thomas.louth@stfc.ac.uk; www.ukatc.stfc.ac.uk

FEA is often the default tool for mechanical engineers looking at the motion of complex structures. However even with a single component that can be modelled in 2D the large design space requires significant computation time to explore. We describe this in chapter 3. A more rapid approach makes use of the fact that even a monolithic structure as studied here can be decomposed into a number of discrete rigid and flexible sections, each of which can be described mathematically. In chapter 2 we make use of existing software tools [4] to exploit this and optimise efficiently. A comparison of the two approaches is provided in chapter 4.

The ability to materially change the positioner tip motion through small changes in geometry is a double-edged sword. It allows the precise path following described in the subsequent chapters, but it also means that small variations in manufactured components will have a noticeable effect on motion. The tools described here can be used to perform sensitivity analysis, as demonstrated in chapter 6.

**UK Astronomy Technology Centre Fibre Positioner Design**

This paper is motivated by the need to optimise a fibre positioner design that is being developed at the UKATC. This design is focused on miniaturization and mass manufacture in order to address the needs of next generation MOS instruments. In particular this design is framed around the specifications of the WST focal plane.

The design is for a theta-r configuration. The two degrees of freedom needed to move around the focal plane are actuated by: an on-axis geared motor providing theta actuation, an off-axis linear actuator which acts parallel to the rotation axis. This actuation is converted to radial motion by a monolithic flexure component incorporating two parallelogram flexures.

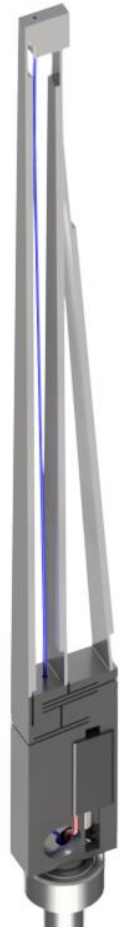

Figure 1. CAD render of the flexure component of the UKATC fibre positioner design

**Simplified Flexure Design**

The analyses presented are based on a parallelogram flexure with the following properties:

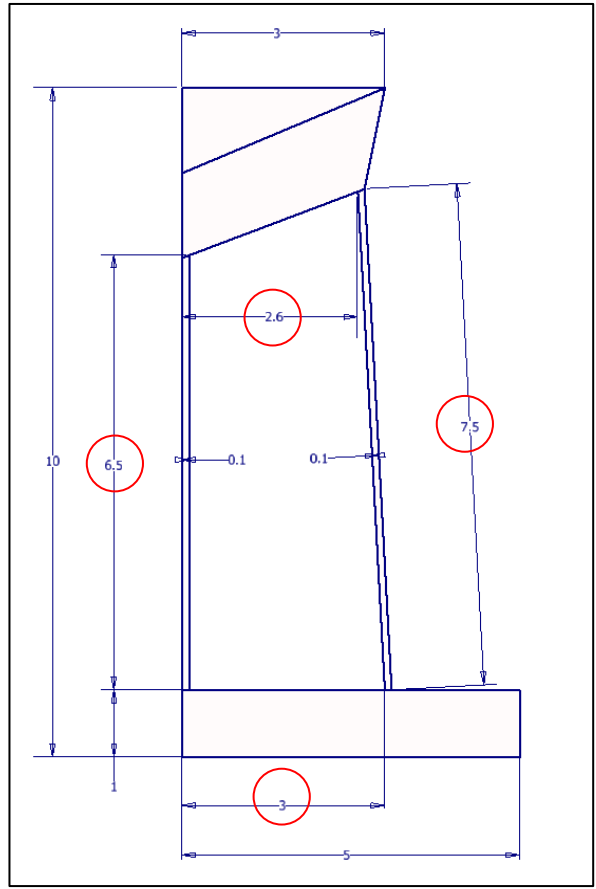

Figure 2. CAD render of the simplified parallelogram flexure, variable dimensions (parameters) highlighted

Table 1: Parameter values

| Parameter | Min | Geom1 | Geom2 | Geom3 | Max |
|---|---|---|---|---|---|
| F1_LEN | 5.0 | 6.5 | 7.0 | 8.0 | 8.0 |
| F2_LEN | 6.5 | 7.5 | 8.5 | 8.0 | 8.5 |
| Bottom_Space | 2.6 | 3.0 | 3.4 | 2.8 | 3.4 |
| Top_Space | 2.6 | 2.6 | 2.6 | 2.8 | 3.0 |

Links 1 and 2 are flexible, and have the following properties.

Table 2: Flexible link properties

| Property | Value |
|---|---|
| In-plane thickness | 0.1 mm |
| Out-of-plane thickness | 1.0 mm |
| Material Stiffness | 70.36 GPa |

Load applied is a single horizontal force of 1N at the midpoint of link 7. Nodes 7 and 8 are fixed to ground.

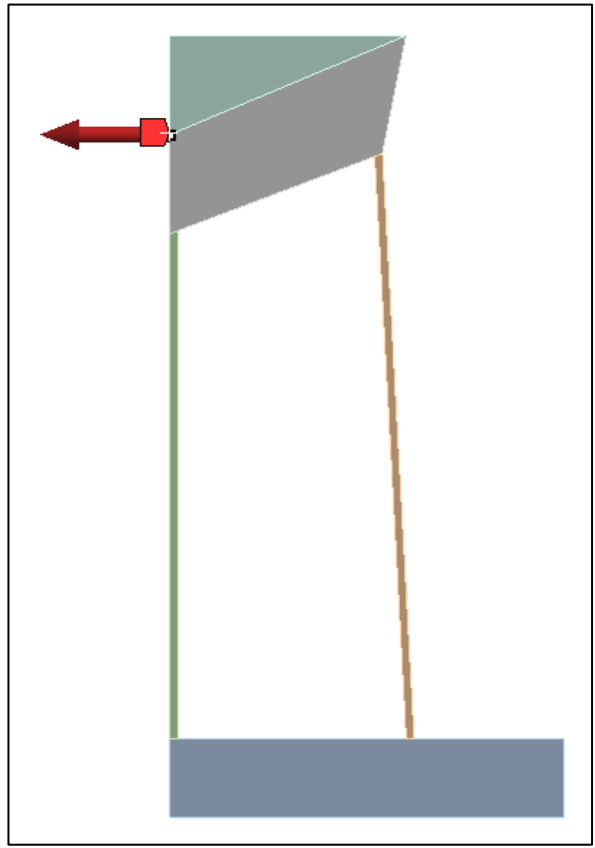

Figure 3. Applied load

**Design Cases**

Three factors are considered which may influence the number of iterations and time taken to solve, and the accuracy of the resulting solution. A number of design cases are used to explore the effect of each factor.

Table 3. Optimisation factors

| Factor | Form | Matlab Implementation | Ansys Implementation | Expected Effect |
| --- | --- | --- | --- | --- |
| Number of active parameters | Which design parameters (dimensions) are considered within the optimisation | A subset of parameters are fixed, others are passed to the optimiser | A subset of parameters are selected within the design optimisation | Positive correlation with number of iterations required |
| Number of degrees of freedom | The geometric complexity of the structure being analysed | Flexible links are divided into a variable number of elements | The mesh density is varied to adjust number of elements | Positive correlation with time to solve each iteration |
| Optimisation threshold | How close to the optimum value the optimiser will search for | fmincon has several relevant parameters described below | The design optimiser has a threshold value for the Objective | Positive correlation with number of iterations required |

Table 4. Design cases

| Case | Number of Parameters | Degrees of Freedom | Optimisation Threshold | Factor Tested |
|---|---|---|---|---|
| Case 1 | 1 | Baseline | Baseline | Number of Parameters |
| Case 2 | 2 | Baseline | Baseline | |
| Case 3 | 3 | Baseline | Baseline | |
| Case 4 | 4 | Baseline | Baseline | |
| Case 5 | 4 | 75% | Baseline | Degrees of Freedom |
| Case 6 | 4 | 125% | Baseline | |
| Case 7 | 4 | 150% | Baseline | |
| Case 8 | 4 | 200% | Baseline | |
| Case 9 | 4 | 250% | Baseline | |
| Case 10 | 4 | Baseline | 300% | Optimiser Threshold |
| Case 11 | 4 | Baseline | 200% | |
| Case 12 | 4 | Baseline | 150% | |
| Case 13 | 4 | Baseline | 25% | |

# 2. ANALYTICAL MATLAB OPTIMISATION

This analysis makes use of the DAS2D Matlab tool developed at Ohio State University by O. Turkkan et al [X]. This allows any 2D flexure geometry to be constructed and analysed. The same approach could be expanded to three dimensions using the DAS3D tool.

**Approach**

The flexure is treated as a series of links between nodes. The unstressed positions of these nodes are either specified explicitly or calculated based on parameter values. Each link is defined as either flexible or rigid.

The DAS2D program takes the geometric definition of the flexure for a given set of parameter values and applied loads and determines the deflection of the structure. Other outputs such as stress are available. The difference between the calculated motion and a predefined ideal path gives the "Objective" value.

Matlab's *fmincon* function is used to find the optimum parameter values that minimise the Objective. Data is publicly available on the functionality of the *fmincon* function and its Interior Point Algorithm so is not reproduced here. Importantly the function is constrained so works within the allowable design space for the parameters. It returns both the parameter values and the value of the Objective at the optimum configuration.

The number of iterations to reach optimum and the total time taken are recorded. This can also provide average time per iteration if desired.

**Setup**

This analysis was performed using Matlab R2019b. Computing hardware used is detailed in the appendices.

The objective is for node 5 to move in a circular arc centred on point (0, 5). Matlab's *norm* function finds the length of the vector between the circle centre and node positions at 10 steps through the motion. The initial distance from the centre to the node (i.e. the target radius) is subtracted from this giving the error at each step. The Objective value is the maximum of the absolute error at any step.

**Analysis Performance**

As hypothesised both the number of iterations and computation time increase with number of parameters considered. The time per iteration is approximately constant. The increase in number of iterations is approximately linear with number of parameters, suggesting that the optimiser is efficient at exploring the design space; the space scales to the power of the number of parameters.

The objective value decreases (improves) as the number of parameters increases in a highly nonlinear way. For very small numbers of parameters it is unsurprising that good optimum geometries cannot be found. Sensitivity analysis (Chapter 6) may be a useful tool for limiting the analysis only to those parameters which strongly affect the outcome.

Table 5. Performance vs number of parameters

| Case | Number of Parameters | Objective Value | Number of Iterations | Computation Time (seconds) | Timer per Iteration (seconds) |
|---|---|---|---|---|---|
| 1 | 1 | 0.00190 | 162 | 72.8 | 0.449 |
| 2 | 2 | 0.00170 | 216 | 98.6 | 0.456 |
| 3 | 3 | 0.00087 | 253 | 118.7 | 0.469 |
| 4 | 4 | 0.00067 | 294 | 136.3 | 0.464 |

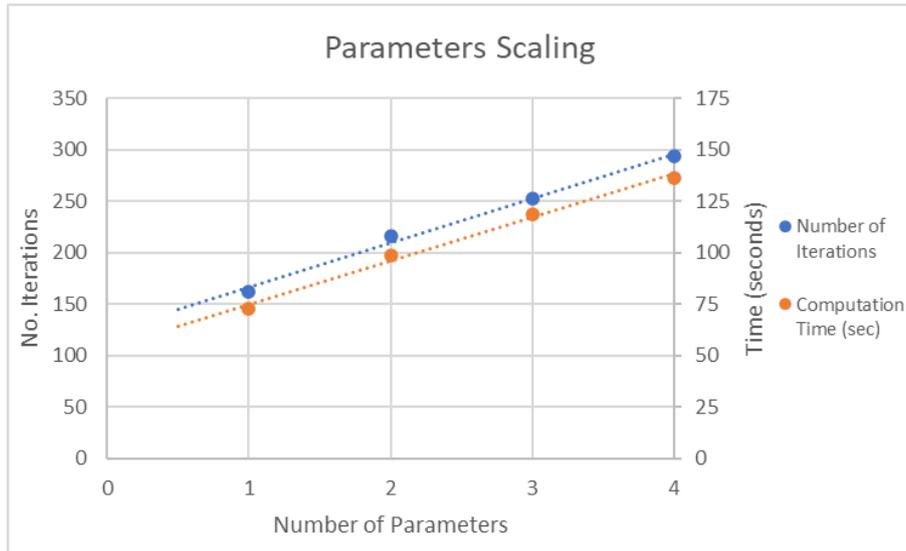

Figure 4. Graph of performance vs number of parameters

As hypothesised the computation time per iteration increases with the number of degrees of freedom of the system; the relationship is linear within the study range. Unexpectedly the number of iterations required to converge increased with DoF, while the resulting objective value decreased giving a better result.

Table 6. Performance vs number of degrees of freedom

| Case | Number of Parameters | Objective Value | Number of Iterations | Computation Time (seconds) | Time per Iteration (sec) |
|---|---|---|---|---|---|
| 5 | 75% | 0.00078 | 287 | 123.9 | 0.432 |
| 4 | 100% | 0.00067 | 294 | 136.3 | 0.464 |
| 6 | 125% | 0.00052 | 288 | 140.8 | 0.489 |
| 7 | 150% | 0.00046 | 346 | 180.9 | 0.523 |
| 8 | 200% | 0.00040 | 354 | 210.8 | 0.595 |
| 9 | 250% | 0.00041 | 315 | 209.7 | 0.666 |

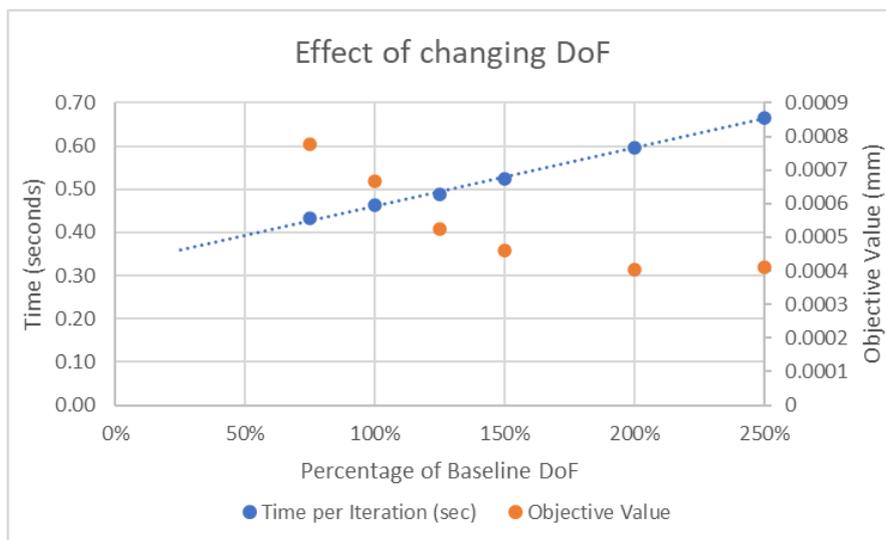

Figure 5. Graph of performance vs degrees of freedom

The effect of changing threshold on analysis performance is more complex. As threshold decreases the number of iterations and computation time approach a cliff. As threshold continues to reduce the optimiser halts for other reasons (step size also has a threshold and approaches zero at the cliff). This cliff is a corollary of convergence in FEA: it demonstrates that the best possible result has been found for the inputs given.

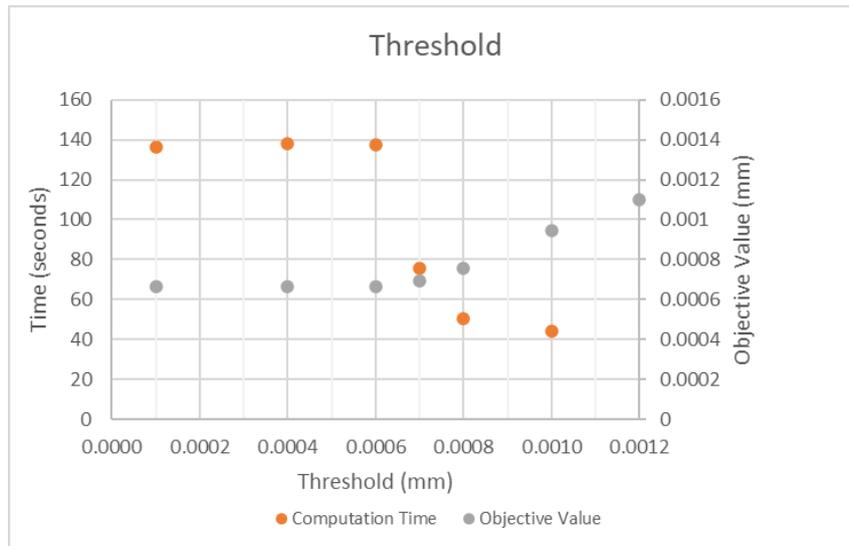

Figure 6. Graph of performance vs threshold. Note cliff at 0.0006mm

**Results Discussion**

The Matlab approach is effective in that it finds geometry which follows the specified path to a tight tolerance, far better than can easily be achieved by hand (0.4 micron objective value vs 10 micron best effort by hand). The compute times for the simple geometry are extremely short, and scaling appears to be broadly linear with complexity. Extrapolating to more complex geometries this should give reasonable (in the order of hours rather than days) compute times for realistic fibre positioner geometries.

If the number of parameters optimised for at any one time can be held low then solve times will be correspondingly short this approach can be a useful design tool for exploring different topologies. Without the use of a GUI the setup of a specific topology (i.e. a specific combination of nodes and links) requires care but is not too time consuming. A comprehensive arrangement of dimensions is required which allows calculation of the unstressed location of each node as the dimensions are varied.

# 3. NUMERICAL ANSYS FEA OPTIMISATION

**Approach**

This is a conventional FEA task modelled with 2D elements. Beam elements are available within Ansys but would not add any fidelity compared to the Matlab approach so are not considered. The entire body is flexible which allows for deformation in the bulk material as well as the flexure components; this is expected to be negligible (see Section 4).

**Setup**

This analysis is based on CAD modelling in Inventor 2021 and FEA in Ansys Workbench 2021. Plugins for other CAD packages are available to give near identical functionality. Optimisation in other FEA packages is available but may not follow a similar workflow. Computing hardware used is detailed in the appendices.

The study is 2D and the mesh is quadrilateral, with 3 elements across the flexible link thickness.

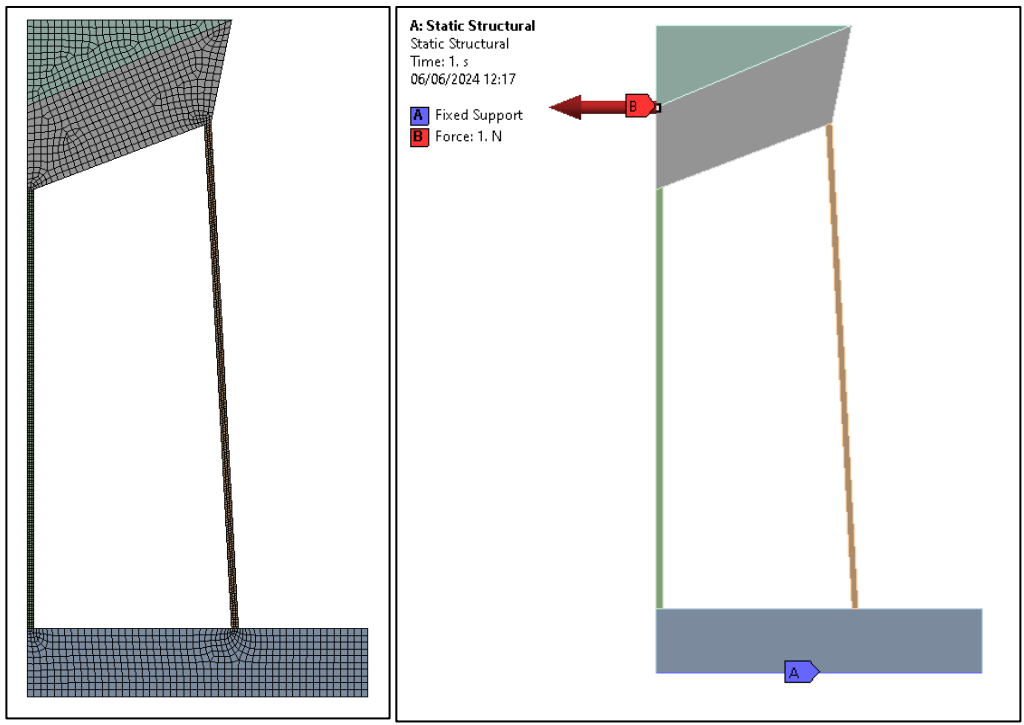

Figure 7. Left: Ansys mesh. Right: Ansys boundary conditions.

Objective function is the difference in radius between the actual path of a node and an optimum circle. This is implemented using a spring connection between a remote point at the node and ground, where the grounded end is fixed at the centre of the optimum circle. The use of a remote point at the node is needed to allow the correct rotation degree of freedom between spring and body. Other physical representations of the objective were considered; this most closely matches that used in the Matlab method. Ansys optimisation has the option to include multiple objectives, but not to combine them in arbitrary ways. For example it can optimise for both node 5 and node 6 motion but cannot optimise only for whichever has the larger absolute value.

The number of iterations (called design points in Ansys WorkBench) is recorded. Total computation time is directly measured.

**Analysis Performance**

Only the effect of changing number of included parameters was found. Solve times were too high to run all design cases and find the scaling factors of changing degrees of freedom and threshold.

Table 7. Performance measures

| Case | Objective Value | Number of Iterations | Computation Time (minutes) |
|---|---|---|---|
| Case 1 (1 parameter) | 0.00105 | 11 | 8 |
| Case 2 (2 parameters) | 0.00060 | 42 | 25 |
| Case 3 (3 parameters) | 0.00051 | 77 | 52 |
| Case 4 (4 parameters) | 0.00093 | 91 | 68 |

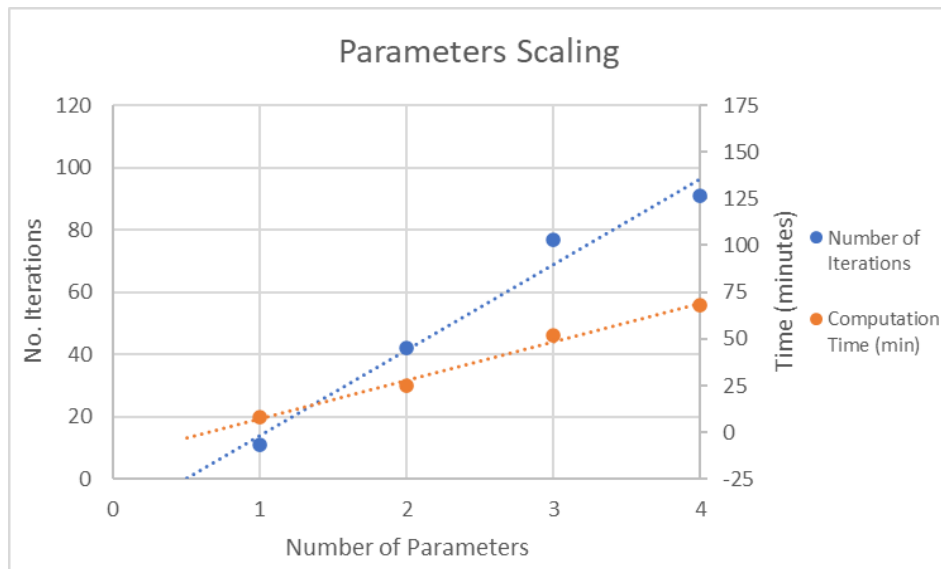

Figure 8. Graph of performance vs number of parameters

**Results Discussion**

The time per iteration for Ansys was high, resulting in excessive time to optimise a single design case. Much of this time is a result of the need for Ansys to connect to the CAD program (Autodesk Inventor in this case) in order to generate the new geometry. This is then brough into Ansys and meshed. A more streamlined FEA approach could avoid some of these steps, perhaps generating the new geometry within a native Ansys program such as SpaceClaim. Equally the mesh could be retained but distorted to the new geometry. Note also that this analysis used large deflection settings within the FEA solver which might not be applicable to all flexure designs.

# 4. COMPARISON OF METHODS

**Correlation**

The two forms of analysis show very similar deformed geometries, though the travel along the path for a given force (i.e. the overall stiffness) is slightly different (by <2%). This difference has not been thoroughly investigated, as the study is concerned primarily with the paths traced by the nodes rather than the distance along those paths.

Node 5 and 6 paths from the two methods have been compared for each of Geom1:3 across the full range of motion. Paths match to <0.2%

Table 8. Correlation

| Geometry | Travel Difference | Node 5 Path Difference | Node 6 Path Difference |
|---|---|---|---|
| Geom1 | 1.1% | 0.02% | 0.04% |
| Geom2 | 1.6% | 0.03% | 0.07% |
| Geom3 | 1.1% | 0.17% | 0.09% |

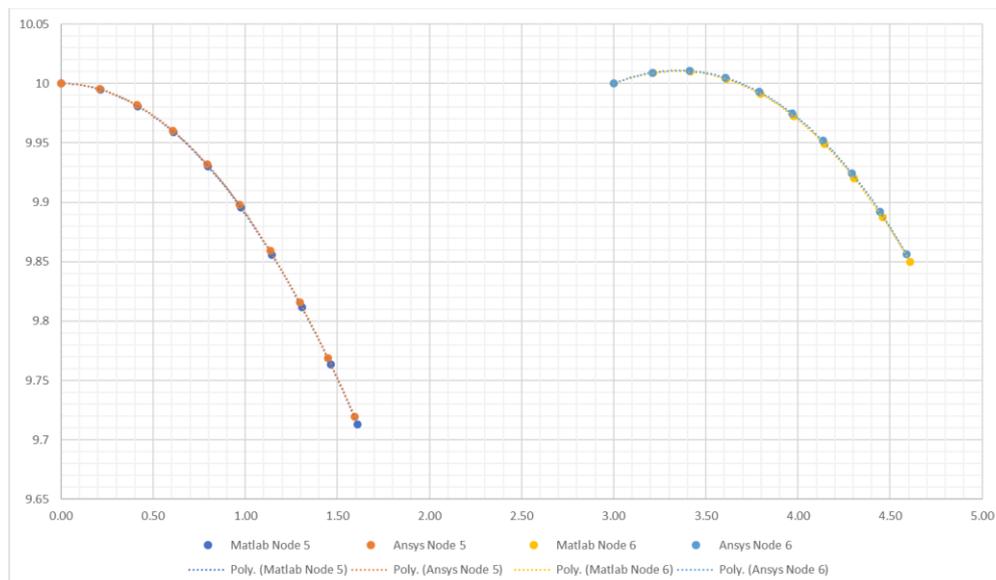

Figure 9. Plot of node 5 and 6 positions through the full motion for Geom1

**Time Taken**

For the geometry studied the Ansys approach is between 10 and 30 times slower than Matlab. This ratio increases with complexity (Ansys time increases faster than Matlab time).

**Optimum Geometry**

The two methods give outputs at similar objective values, however the corresponding geometry does not always match. This may be a result of the two methods finding different local minima.

**Scaling**

For both approaches the computation time appears to linearly increase with number of parameters. Given that the design space increases to the power of the number of parameters this demonstrates that the optimiser algorithms are rather efficient.

# 5. APPLICABILITY TO UKATC FIBRE POSITIONER

**Approach**

The approach follows the Matlab method, and the initial and optimum parameter values are checked using Ansys. Full optimisation using Ansys has been discounted due to the excessive computation time predicted.

**Setup**

The geometry is divided into 23 links between 22 nodes. Nodes 21 and 22 are fixed to ground, and an upwards force of 0.2N is applied to the midpoint of link 19. Links 1, 2, 3, 4, 5 are flexible with in-plane thickness 0.1mm; links 5, 8, 9, 12 are flexible with in-plane thickness 0.5mm; all other links are rigid.

The objective is for nodes 19 and 20 to move in a circular arc centred on point (0, 10 000), representing a telecentric, concave focal surface with focal length 10m. The objective value is calculated as described in Chapter 2.

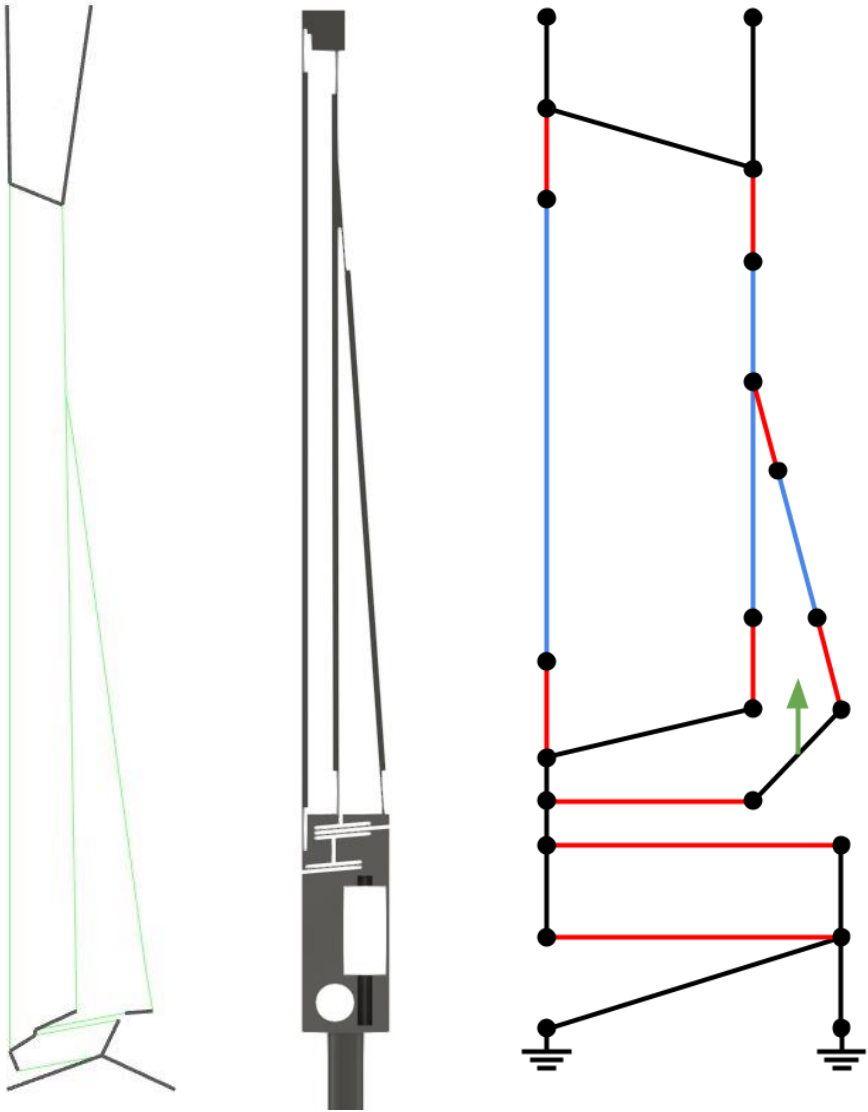

Figure 10. Matlab plot of unstressed geometry, CAD render, topological representation of the positioner

**Analysis Performance**

The optimum Objective value found was higher than for the simplified flexure at ~12 microns; this suggests that the arrangement of flexure elements or the allowable range of the dimensions does not produce a completely optimised positioner.

Table 9. Performance measures

| Objective Value | Number of Iterations | Computation Time (minutes) |
|---|---|---|
| 0.0123 | 1872 | 150 |

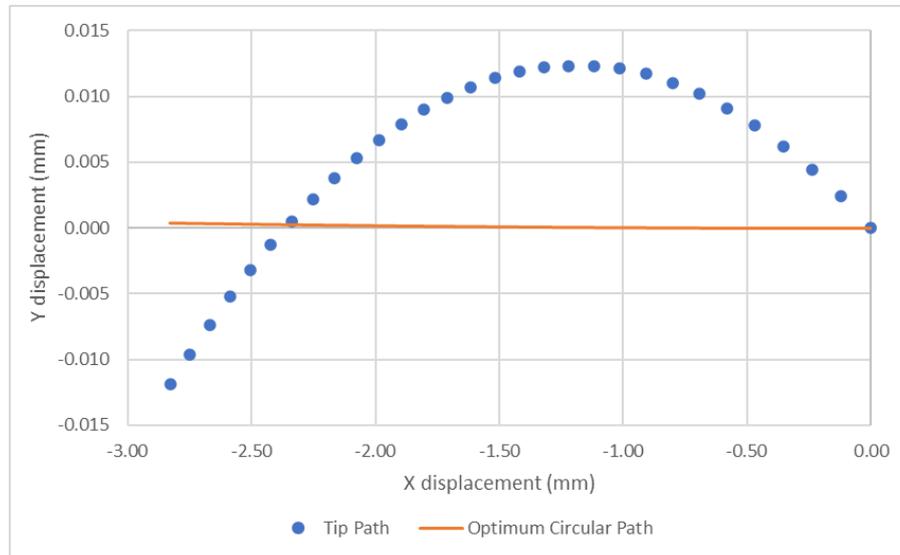

Figure 11. Motion of the positioner tip for the optimum geometry.

**Results Discussion**

It is clear that the combination of flexure elements used in the positioner concept does not result in a path that closely approximates the curved focal plane. The tip path seen is a combination of downwards curve from the parallelogram and an approximately linear upwards correction from the relieving flexures. To achieve a better path it will be necessary to find an arrangement of relieving flexures that has a significantly non-linear effect.

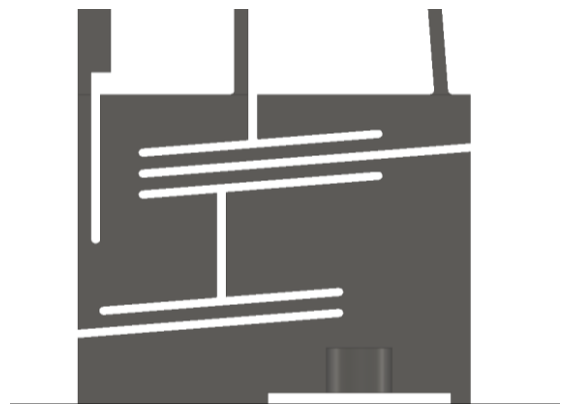

Figure 12. Relieving flexures detail

## 6. SENSITIVITY ANALYSIS

**Approach**

Sensitivity analysis was performed in Matlab using the same code as the optimisation, but only varying one parameter at a time; all other parameter values are fixed at the optimum result. In this way the change in objective value for a given change in each input parameter can be found.

This serves two purposes. Where the objective is found to be insensitive to a specific parameter this can be discounted from further optimisation. Given the sensitivity of the analysis to the number of input parameters this can reduce optimisation time significantly. A possible workflow is to do a coarse optimisation with a loose objective threshold, then a sensitivity analysis around this point, fix any insensitive dimensions at reasonable values, then do a fine optimisation with fewer active parameters.

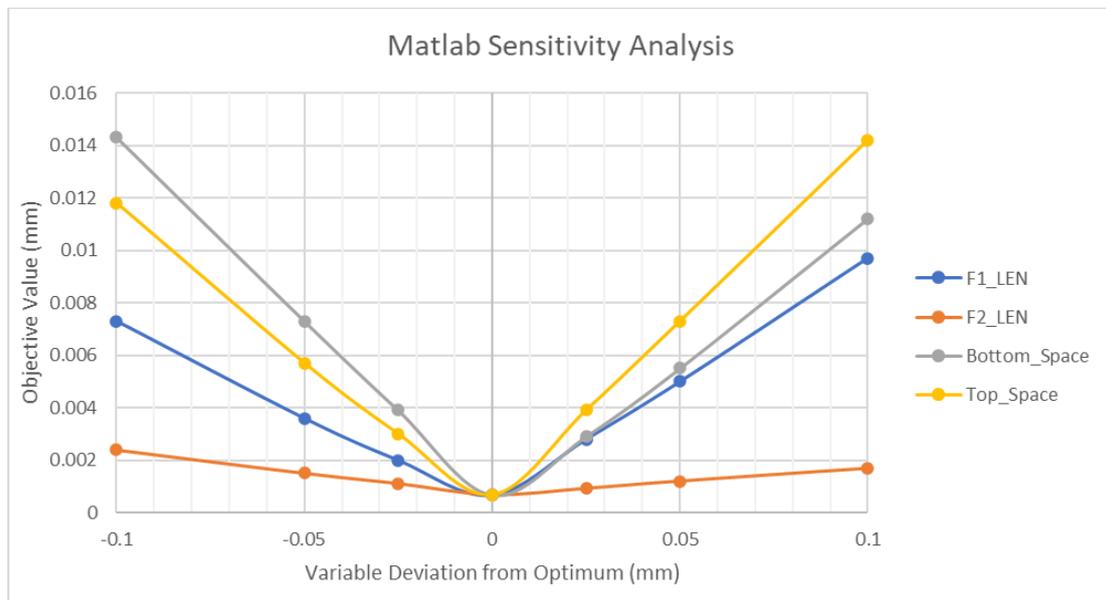

Figure 13. Sensitivity of the simplified geometry to each parameter. The objective has low sensitivity to F2_LEN, so this might be ignored in future optimisation.

## 7. CONCLUSIONS

These analyses have been performed on early concept designs; other design constraints may impose changes that necessitate new optimisation. As such, the analysis should be efficient to set up and run. Both approaches described here are efficient to set up, requiring only that the geometry be simplified into discrete elements and two dimensions respectively. For FEA design optimisation a robust CAD model is needed to prevent the geometry falling over as dimension changes are automatically implemented. This may be hard to maintain as geometry becomes more complex; one option is to maintain a parallel simplified model that matches the relevant features of the full model.

A shortcoming of this study is that the optimum paths considered are in all cases circular arcs. No method was found in Ansys to automatically determine the distance between a deformed nodal position and an arbitrary curve (for example a high order polynomial); no doubt this could be achieved through the use of APDL scripting. The Matlab method is more adaptable in this regard in that there is much more freedom in creating objective functions. One option to overcome this limitation of the Ansys method would be to use software such as Matlab for the optimisation while using Ansys FEA to provide the deformed geometry. There might also be value in using Ansys beam elements although this would reduce the fidelity of the Ansys geometry almost exactly to the same level as in the Matlab method.

For real fibre positioners it is necessary to optimise for tilt as well as focus, and it may be important to include other considerations such as minimizing stiffness. It is not trivial to define good objective functions for these combined cases. Do you optimise to minimise whichever error is larger, or do you have a weighted contribution for each metric?

## 8. APPENDICES

**Computing Hardware**

Windows 11

CPU 12 Core Intel i9-10920X @ 3.50 GHz

Memory 64GB 2666MHz 4 slots

Disk SSD PC SN730 NVMe WDC 1024GB

**Matlab Code**

The setup and optimisation code is available at:

**https://drive.google.com/file/d/1Ja1dd7Lhc33OgcGRy4Aw7P7ZM1lud0EO/view?usp=sharing**

**Matlab Tricks**

If the DAS2D routine is unstable then increasing the number of increments can help. The simplified geometry solved with 10 increments, UKATC flexure required 30 increments.

Number of iterations to find optimum is also affected by the available range for each parameter. Think about acceptable maximum and minimum values to constrain this range.

## ACKNOWLEDGEMENTS

This study made extensive use of the Matlab tools produced by O Turkkan et al at Ohio State University, and made available at:
**http://www.compliantanalysis.com**